\documentclass[aps,preprint]{revtex4}%
\usepackage{amsfonts}
\usepackage{amsmath}
\usepackage{amssymb}
\usepackage{graphicx}%
\providecommand{\U}[1]{\protect\rule{.1in}{.1in}}
\begin{document}
\title{Variance Control in Weak Value Measurement Pointers}
\author{A. D. Parks and J. E. Gray}
\affiliation{\textit{Electromagnetic and Sensor Systems Department, Naval Surface Warfare
Center, Dahlgren, VA 22448, USA}}

\pacs{03.65.-w, 03.65.Ca, 03.65.Ta, 06.20.Dk}

\begin{abstract}
The variance of an arbitrary pointer observable is considered for the general
case that a complex weak value is measured using a complex valued pointer
state. For the typical cases where the pointer observable is either its
position or momentum, the associated expressions for the pointer's variance
after the measurement contain a term proportional to the product of the weak
value's imaginary part with the rate of change of the third central moment of
position relative to the initial pointer state just prior to the time of the
measurement interaction when position is the observable - or with the initial
pointer state's third central moment of momentum when momentum is the
observable. These terms provide a means for controlling pointer position and
momentum variance and identify control conditions which - when satisfied - can
yield variances that are smaller after the measurement than they were before
the measurement. Measurement sensitivities which are useful for estimating
weak value measurement accuracies are also briefly discussed.

\end{abstract}
\date{April 11, 2011}
\startpage{1}
\endpage{102}
\maketitle

\section{Introduction}

The weak value $A_{w}$ of a quantum mechanical observable $A$ was introduced
by Aharonov \textit{et al} \cite{A1, A2, A3} a quarter century ago. This
quantity is the statistical result of a standard measurement procedure
performed upon a pre- and post-selected (PPS) ensemble of quantum systems when
the interaction between the measurement apparatus and each system is
sufficiently weak, i.e. when it is a weak measurement. Unlike a standard
strong measurement of $A$ which significantly disturbs the measured system
(i.e. it "collapses" the wavefunction), a weak measurement of $A$ for a PPS
system does not appreciably disturb the quantum system and yields $A_{w}$ as
the observable's measured value. The peculiar nature of the virtually
undisturbed quantum reality that exists between the boundaries defined by the
PPS states is revealed in the eccentric characteristics of $A_{w}$, namely
that $A_{w}$ can be complex valued and that the real part $\operatorname{Re}%
A_{w}$ of $A_{w}$ can lie far outside the eigenvalue spectral limits of
$\widehat{A}$. While the interpretation of weak values remains somewhat
controversial, experiments have verified several of the interesting unusual
properties predicted by weak value theory \cite{R, P, RL, W, H, Y, D}.

The pointer of a measurement apparatus is fundamental to the theory of quantum
measurement because the values of measured observables are determined from the
pointer's properties (e.g. from the pointer's mean position). Understanding
these properties has become more important in recent years - in large part due
to the increased interest in the theory of weak measurements and weak value
theory. The properties of pointers associated with weak value measurements
have been studied - for example - by Johansen \cite{J}, Aharonov and Botero
\cite{AB}, Jozsa \cite{Jo}, Di Lorenzo and Egues \cite{DE}, and Cho \textit{et
al }\cite{C}.

The purpose of this paper is to extend Jozsa's work \cite{Jo} to obtain the
general expression for the variance associated with an arbitrary pointer
observable when a complex valued pointer state is used to measure a complex
weak value $A_{w}$. For the typical cases where position or momentum are the
pointer observables, the associated expressions each contain a variance
control term. This term is proportional to the product of the imaginary part
$\operatorname{Im}A_{w}$ of $A_{w}$ with the rate of change of the third
central moment of position relative to the initial pointer state just prior to
measurement when the observable is position - or with the initial pointer
state's third central moment of momentum when momentum is the observable.
Control conditions associated with these terms are identified which - if
satisfied - can yield pointer position and momentum variances after a
measurement that are smaller than they were prior to the measurement. These
results are used to briefly discuss sensitivities associated with weak value measurements.

\section{Weak Measurements and Weak Values}

For the reader's convenience, this section provides a brief review of weak
measurement and weak value theory. For additional details the reader is
invited to consult references \cite{A1, A2, A3, A4}.

Weak measurements arise in the von Neumann description of a quantum
measurement at time $t_{0}$ of a time-independent observable $A$ that
describes a quantum system in an initial fixed pre-selected state $\left\vert
\psi_{i}\right\rangle =%
%TCIMACRO{\dsum \nolimits_{J}}%
%BeginExpansion
{\displaystyle\sum\nolimits_{J}}
%EndExpansion
c_{j}\left\vert a_{j}\right\rangle $ at $t_{0}$, where the set $J$ indexes the
eigenstates $\left\vert a_{j}\right\rangle $ of $\widehat{A}$. In this
description, the Hamiltonian for the interaction between the measurement
apparatus and the quantum system is
\[
\widehat{H}=\gamma(t)\widehat{A}\widehat{p}.
\]
Here $\gamma(t)=\gamma\delta(t-t_{0})$ defines the strength of the
measurement's impulsive coupling interaction at $t_{0}$ and $\widehat{p}$ is
the momentum operator for the pointer of the measurement apparatus which is in
the initial normalized state $\left\vert \phi\right\rangle $. Let $\widehat
{q}$ be the pointer's position operator that is conjugate to $\widehat{p}$.

Prior to the measurement the pre-selected system and the pointer are in the
tensor product state $\left\vert \psi_{i}\right\rangle \left\vert
\phi\right\rangle $. Immediately following the interaction the combined system
is in the state%
\[
\left\vert \Phi\right\rangle =e^{-\frac{i}{\hbar}\int\widehat{H}dt}\left\vert
\psi_{i}\right\rangle \left\vert \phi\right\rangle =%
%TCIMACRO{\dsum \nolimits_{J}}%
%BeginExpansion
{\displaystyle\sum\nolimits_{J}}
%EndExpansion
c_{j}e^{-\frac{i}{\hbar}\gamma a_{j}\widehat{p}}\left\vert a_{j}\right\rangle
\left\vert \phi\right\rangle ,
\]
where use has been made of the fact that $\int\widehat{H}dt=\gamma\widehat
{A}\widehat{p}$. The exponential factor in this equation is the translation
operator $\widehat{S}\left(  \gamma a_{j}\right)  $ for $\left\vert
\phi\right\rangle $ in its $q$-representation. It is defined by the action
$\left\langle q\right\vert \widehat{S}\left(  \gamma a_{j}\right)  \left\vert
\phi\right\rangle $ which translates the pointer's wavefunction over a
distance $\gamma a_{j}$ parallel to the $q$-axis. The $q$-representation of
the combined system and pointer state is%
\[
\left\langle q\right\vert \left.  \Phi\right\rangle =%
%TCIMACRO{\dsum \nolimits_{J}}%
%BeginExpansion
{\displaystyle\sum\nolimits_{J}}
%EndExpansion
c_{j}\left\langle q\right\vert \widehat{S}\left(  \gamma a_{j}\right)
\left\vert \phi\right\rangle \left\vert a_{j}\right\rangle .
\]

When the measurement interaction is strong, the quantum system is appreciably
disturbed and its state "collapses" with probability $\left\vert
c_{n}\right\vert ^{2}$ to an eigenstate $\left\vert a_{n}\right\rangle $
leaving the pointer in the state $\left\langle q\right\vert \widehat{S}\left(
\gamma a_{n}\right)  \left\vert \phi\right\rangle $. Strong measurements of an
ensemble of identically prepared systems yield $\gamma\left\langle
A\right\rangle \equiv\gamma\left\langle \psi_{i}\right\vert \widehat
{A}\left\vert \psi_{i}\right\rangle $ as the centroid of the associated
pointer probability distribution with $\left\langle A\right\rangle $ as the
measured value of $\widehat{A}$.

A \textit{weak measurement }of $\widehat{A}$ occurs when the interaction
strength $\gamma$ is sufficiently small so that the system is essentially
undisturbed and the uncertainty $\Delta q$ is much larger than $\widehat{A}$'s
eigenvalue separation. In this case, the pointer distribution is the
superposition of broad overlapping $\left\vert \left\langle q\right\vert
\widehat{S}\left(  \gamma a_{j}\right)  \left\vert \phi\right\rangle
\right\vert ^{2}$ terms. Although a single measurement provides little
information about $\widehat{A}$, many repetitions allow the centroid of the
distribution to be determined to any desired accuracy.

If a system is post-selected after a weak measurement is performed, then the
resulting pointer state is%
\[
\left\vert \Psi\right\rangle \equiv\left\langle \psi_{f}\right\vert \left.
\Phi\right\rangle =%
%TCIMACRO{\dsum \nolimits_{J}}%
%BeginExpansion
{\displaystyle\sum\nolimits_{J}}
%EndExpansion
c_{j}^{\prime\ast}c_{j}\widehat{S}\left(  \gamma a_{j}\right)  \left\vert
\phi\right\rangle ,
\]
where $\left\vert \psi_{f}\right\rangle =%
%TCIMACRO{\dsum \nolimits_{J}}%
%BeginExpansion
{\displaystyle\sum\nolimits_{J}}
%EndExpansion
c_{j}^{\prime}\left\vert a_{j}\right\rangle $, $\left\langle \psi
_{f}\right\vert \left.  \psi_{i}\right\rangle \neq0$, is the post-selected
state at $t_{0}$. Since%
\[
\widehat{S}\left(  \gamma a_{j}\right)  =%
%TCIMACRO{\dsum \limits_{m=0}^{\infty}}%
%BeginExpansion
{\displaystyle\sum\limits_{m=0}^{\infty}}
%EndExpansion
\frac{\left(  -i\gamma a_{j}\widehat{p}/\hbar\right)  ^{m}}{m!},
\]
then%
\[
\left\vert \Psi\right\rangle =%
%TCIMACRO{\dsum \nolimits_{J}}%
%BeginExpansion
{\displaystyle\sum\nolimits_{J}}
%EndExpansion
c_{j}^{\prime\ast}c_{j}\left\{  1-\frac{i}{\hbar}\gamma A_{w}\widehat{p}+%
%TCIMACRO{\dsum \limits_{m=2}^{\infty}}%
%BeginExpansion
{\displaystyle\sum\limits_{m=2}^{\infty}}
%EndExpansion
\frac{\left(  -i\gamma\widehat{p}/\hbar\right)  ^{m}}{m!}\left(  A^{m}\right)
_{w}\right\}  \left\vert \phi\right\rangle \approx\left\{
%TCIMACRO{\dsum \nolimits_{J}}%
%BeginExpansion
{\displaystyle\sum\nolimits_{J}}
%EndExpansion
c_{j}^{\prime\ast}c_{j}\right\}  e^{-\frac{i}{\hbar}\gamma A_{w}\widehat{p}%
}\left\vert \phi\right\rangle
\]
in which case%
\[
\left\vert \Psi\right\rangle \approx\left\langle \psi_{f}\right\vert \left.
\psi_{i}\right\rangle \widehat{S}\left(  \gamma A_{w}\right)  \left\vert
\phi\right\rangle .
\]
Here%
\[
\left(  A^{m}\right)  _{w}=\frac{%
%TCIMACRO{\dsum \nolimits_{J}}%
%BeginExpansion
{\displaystyle\sum\nolimits_{J}}
%EndExpansion
c_{j}^{\prime\ast}c_{j}a_{j}^{m}}{%
%TCIMACRO{\dsum \nolimits_{J}}%
%BeginExpansion
{\displaystyle\sum\nolimits_{J}}
%EndExpansion
c_{j}^{\prime\ast}c_{j}}=\frac{\left\langle \psi_{f}\right\vert \widehat
{A}^{m}\left\vert \psi_{i}\right\rangle }{\left\langle \psi_{f}\right\vert
\left.  \psi_{i}\right\rangle },
\]
with the \textit{weak value} $A_{w}$ of $\widehat{A}$ defined by%
\begin{equation}
A_{w}\equiv\left(  A^{1}\right)  _{w}=\frac{\left\langle \psi_{f}\right\vert
\widehat{A}\left\vert \psi_{i}\right\rangle }{\left\langle \psi_{f}\right\vert
\left.  \psi_{i}\right\rangle }. \label{3}%
\end{equation}
From this expression it is obvious that $A_{w}$ is - in general - a complex
valued quantity that can be calculated directly from theory and that when the
PPS states are nearly orthogonal $\operatorname{Re}A_{w}$ can lie far outside
$\widehat{A}$'s eigenvalue spectral limits.

For the general case where both $A_{w}$ and $\phi\left(  q\right)  $ are
complex valued, the mean pointer position and momentum after a measurement are
given by \cite{Jo}%
\begin{equation}
\left\langle \Psi\right\vert \widehat{q}\left\vert \Psi\right\rangle
=\left\langle \phi\right\vert \widehat{q}\left\vert \phi\right\rangle
+\gamma\operatorname{Re}A_{w}+\left(  \frac{\gamma}{\hbar}\right)
\operatorname{Im}A_{w}\left(  m\frac{d\Delta_{\phi}^{2}q}{dt}\right)
\label{2}%
\end{equation}
and
\begin{equation}
\left\langle \Psi\right\vert \widehat{p}\left\vert \Psi\right\rangle
=\left\langle \phi\right\vert \widehat{p}\left\vert \phi\right\rangle
+2\left(  \frac{\gamma}{\hbar}\right)  \operatorname{Im}A_{w}\left(
\Delta_{\phi}^{2}p\right)  , \label{2A}%
\end{equation}
respectively. Here $m$ is the mass of the pointer, $\Delta_{\phi}^{2}p$ is the
pointer's initial momentum variance, and the time derivative of $\Delta_{\phi
}^{2}q$ is the rate of change of the initial pointer position variance just
prior to $t_{0}$.

\section{Pointer Variance}

The mean value of an arbitrary pointer observable $M$ after a measurement of
$A_{w}$ is \cite{Jo}%
\begin{align}
\left\langle \Psi\right\vert \widehat{M}\left\vert \Psi\right\rangle  &
=\left\langle \phi\right\vert \widehat{M}\left\vert \phi\right\rangle
-i\left(  \frac{\gamma}{\hbar}\right)  \operatorname{Re}A_{w}\left\langle
\phi\right\vert \left[  \widehat{M},\widehat{p}\right]  \left\vert
\phi\right\rangle +\label{4}\\
&  \left(  \frac{\gamma}{\hbar}\right)  \operatorname{Im}A_{w}\left(
\left\langle \phi\right\vert \left\{  \widehat{M},\widehat{p}\right\}
\left\vert \phi\right\rangle -2\left\langle \phi\right\vert \widehat
{M}\left\vert \phi\right\rangle \left\langle \phi\right\vert \widehat
{p}\left\vert \phi\right\rangle \right)  ,\nonumber
\end{align}
where $\left\{  \widehat{M},\widehat{p}\right\}  =\widehat{M}\widehat
{p}+\widehat{p}\widehat{M}$. Note that eq.(\ref{4}) reduces to eq.(\ref{2A})
when $\widehat{M}=\widehat{p}$ and that it is also in complete agreement with
eq.(\ref{2}) when $\widehat{M}=\widehat{q}$ since $\left[  \widehat
{q},\widehat{p}\right]  =i\hbar$ and the equations of motion for $\left\langle
\phi\right\vert \widehat{q}\left\vert \phi\right\rangle $ and $\left\langle
\phi\right\vert \widehat{q}^{2}\left\vert \phi\right\rangle $ yield%
\begin{equation}
\left\langle \phi\right\vert \left\{  \widehat{q},\widehat{p}\right\}
\left\vert \phi\right\rangle =m\frac{d\left\langle \phi\right\vert \widehat
{q}^{2}\left\vert \phi\right\rangle }{dt} \label{8}%
\end{equation}
and%
\begin{equation}
2\left\langle \phi\right\vert \widehat{q}\left\vert \phi\right\rangle
\left\langle \phi\right\vert \widehat{p}\left\vert \phi\right\rangle
=m\frac{d\left\langle \phi\right\vert \widehat{q}\left\vert \phi\right\rangle
^{2}}{dt}. \label{9}%
\end{equation}
Here the time derivatives are rates of change of the corresponding quantities
just prior to the interaction time $t_{0}$.

The pointer variance for $M$ is easily determined from eq.(\ref{4}) by
subtracting its square from the expression obtained from eq.(\ref{4}) when
$\widehat{M}$ is replaced by $\widehat{M}^{2}$. Retaining terms through first
order in $\left(  \frac{\gamma}{\hbar}\right)  $ yields the following result:
\begin{equation}
\Delta_{\Psi}^{2}M=\Delta_{\phi}^{2}M-i\left(  \frac{\gamma}{\hbar}\right)
\operatorname{Re}A_{w}\mathcal{F}\left(  \widehat{M}\right)  +\left(
\frac{\gamma}{\hbar}\right)  \operatorname{Im}A_{w}\mathcal{G}\left(
\widehat{M}\right)  . \label{5}%
\end{equation}
Here $\Delta_{\phi}^{2}M$ and $\Delta_{\Psi}^{2}M$ are the initial and final
variances, respectively,
\[
\mathcal{F}\left(  \widehat{M}\right)  \equiv\left\langle \phi\right\vert
\left[  \widehat{M}^{2},\widehat{p}\right]  \left\vert \phi\right\rangle
-2\left\langle \phi\right\vert \widehat{M}\left\vert \phi\right\rangle
\left\langle \phi\right\vert \left[  \widehat{M},\widehat{p}\right]
\left\vert \phi\right\rangle ,
\]
and%
\begin{equation}
\mathcal{G}\left(  \widehat{M}\right)  \equiv\left\langle \phi\right\vert
\left\{  \widehat{M}^{2},\widehat{p}\right\}  \left\vert \phi\right\rangle
-2\left\langle \phi\right\vert \widehat{M}\left\vert \phi\right\rangle
\left\langle \phi\right\vert \left\{  \widehat{M},\widehat{p}\right\}
\left\vert \phi\right\rangle -2\left\langle \phi\right\vert \widehat
{p}\left\vert \phi\right\rangle \left(  \Delta_{\phi}^{2}M-\left\langle
\phi\right\vert \widehat{M}\left\vert \phi\right\rangle ^{2}\right)  .
\label{7}%
\end{equation}
As anticipated from eq.(\ref{4}), eq.(\ref{5}) clearly shows that for such a
measurement the pointer variance associated with an arbitrary pointer
observable is also generally effected by both the real and imaginary parts of
the weak value.

However, for the typical cases of interest where $\widehat{M}=\widehat{q}$ or
$\widehat{M}=\widehat{p}$, the pointer's variance is independent of
$\operatorname{Re}A_{w}$ because%

\[
\mathcal{F}\left(  \widehat{q}\right)  =0=\mathcal{F}\left(  \widehat
{p}\right)  .
\]
Here use has been made of the facts that $\left[  \widehat{q},\widehat
{p}\right]  =i\hbar$ and $\left[  \widehat{q}^{2},\widehat{p}\right]
=2i\hbar\widehat{q}$. Consequently, for these cases eq.(\ref{5}) can be
written as%
\begin{equation}
\Delta_{\Psi}^{2}M=\Delta_{\phi}^{2}M+\left(  \frac{\gamma}{\hbar}\right)
\operatorname{Im}A_{w}\mathcal{G}\left(  \widehat{M}\right)  ,\text{ }M=q,p.
\label{11}%
\end{equation}

Now consider $\mathcal{G}\left(  \widehat{M}\right)  $ in more detail. When
$\widehat{M}=\widehat{q}$, then eq.(\ref{7}) becomes
\begin{equation}
\mathcal{G}\left(  \widehat{q}\right)  =\left\langle \phi\right\vert \left\{
\widehat{q}^{2},\widehat{p}\right\}  \left\vert \phi\right\rangle
-2\left\langle \phi\right\vert \widehat{q}\left\vert \phi\right\rangle
\left\langle \phi\right\vert \left\{  \widehat{q},\widehat{p}\right\}
\left\vert \phi\right\rangle -2\left\langle \phi\right\vert \widehat
{p}\left\vert \phi\right\rangle (\Delta_{\phi}^{2}q-\left\langle
\phi\right\vert \widehat{q}\left\vert \phi\right\rangle ^{2}). \label{10}%
\end{equation}
From the equation of motion for $\left\langle \phi\right\vert \widehat{q}%
^{3}\left\vert \phi\right\rangle $ it is found that%
\[
\frac{d\left\langle \phi\right\vert \widehat{q}^{3}\left\vert \phi
\right\rangle }{dt}=-\frac{i}{\hbar}\left\langle \phi\right\vert \left[
\widehat{q}^{3},\widehat{H}\right]  \left\vert \phi\right\rangle =-\frac
{i}{2m\hbar}\left\langle \phi\right\vert \left[  \widehat{q}^{3},\widehat
{p}^{2}\right]  \left\vert \phi\right\rangle =\frac{3}{2m}\left\langle
\phi\right\vert \left\{  \widehat{q}^{2},\widehat{p}\right\}  \left\vert
\phi\right\rangle ,
\]
where $\widehat{H}=\frac{\widehat{p}^{2}}{2m}+V(\widehat{q})$ is the pointer's
Hamiltonian operator. Applying this result - along with eqs.(\ref{8}) and
(\ref{9}) - to eq.(\ref{10}) yields
\[
\mathcal{G}\left(  \widehat{q}\right)  =\frac{2m}{3}\frac{dq_{3}}{dt},
\]
so that eq.(\ref{11}) can be compactly written as%
\[
\Delta_{\Psi}^{2}q=\Delta_{\phi}^{2}q+\frac{2\gamma m}{3\hbar}%
\operatorname{Im}A_{w}\left(  \frac{dq_{3}}{dt}\right)  .
\]
Here $q_{3}\equiv\left\langle \phi\right\vert \left(  \widehat{q}-\left\langle
\phi\right\vert \widehat{q}\left\vert \phi\right\rangle \right)
^{3}\left\vert \phi\right\rangle $ is the third central moment of $\widehat
{q}$ relative to the initial pointer state and its time derivative is the rate
of change of $q_{3}$ just prior to $t_{0}$.

When $\widehat{M}=\widehat{p}$, then eq.(\ref{7}) becomes%
\[
\mathcal{G}\left(  \widehat{p}\right)  =2\left[  \left\langle \phi\right\vert
\widehat{p}^{3}\left\vert \phi\right\rangle -3\left\langle \phi\right\vert
\widehat{p}\left\vert \phi\right\rangle \left\langle \phi\right\vert
\widehat{p}^{2}\left\vert \phi\right\rangle +2\left\langle \phi\right\vert
\widehat{p}\left\vert \phi\right\rangle ^{3}\right]  =2p_{3},
\]
where $p_{3}\equiv\left\langle \phi\right\vert \left(  \widehat{p}%
-\left\langle \phi\right\vert \widehat{p}\left\vert \phi\right\rangle \right)
^{3}\left\vert \phi\right\rangle $ is the third central moment of $\widehat
{p}$ relative to the pointer's initial state, and eq.(\ref{11}) assumes the
form%
\[
\Delta_{\Psi}^{2}p=\Delta_{\phi}^{2}p+2\left(  \frac{\gamma}{\hbar}\right)
\operatorname{Im}A_{w}\left(  p_{3}\right)  .
\]
\ 

The quantities $q_{3}$ and $p_{3}$ provide measures of the skewness of the
pointer position and momentum probability distribution profiles. If the
pointer position profile's skewness is fixed, then $\frac{dq_{3}}{dt}=0$ and
$\Delta_{\Psi}^{2}q=\Delta_{\phi}^{2}q$. Otherwise, $\Delta_{\Psi}^{2}q$ can
be manipulated through the judicious selection of the control term
$\operatorname{Im}A_{w}\left(  \frac{dq_{3}}{dt}\right)  $. In particular,
observe that $0<\Delta_{\Psi}^{2}q\leq\Delta_{\phi}^{2}q$ when this control
term satisfies the inequality
\begin{equation}
\text{ }-\left(  \frac{3\hbar}{2\gamma m}\right)  \Delta_{\phi}^{2}%
q<\operatorname{Im}A_{w}\left(  \frac{dq_{3}}{dt}\right)  \leq0. \label{SQ}%
\end{equation}
Similarly, the control term $\operatorname{Im}A_{w}\left(  p_{3}\right)  $ can
be used to manipulate $\Delta_{\Psi}^{2}p$ when it satisfies the inequality%
\begin{equation}
-\left(  \frac{\hbar}{2\gamma}\right)  \Delta_{\phi}^{2}p<\operatorname{Im}%
A_{w}\left(  p_{3}\right)  \leq0. \label{SQA}%
\end{equation}
Thus, when measuring complex weak values the final pointer position (momentum)
variance can be made smaller than its initial value by choosing
$\operatorname{Im}A_{w}$ or $\frac{dq_{3}}{dt}$ ($p_{3}$) so that condition
(\ref{SQ}) ((\ref{SQA})) is satisfied.

\section{Closing Remarks}

Because of the growing interest in the practical application of weak values,
estimating their measurement sensitivities has also become important from both
the experimental and device engineering perspectives. Applying the calculus of
error propagation to the above results defines the measurement sensitivities
$\delta_{q}\operatorname{Re}A_{w}$ and $\delta_{q}\operatorname{Im}A_{w}$ for
determining $\operatorname{Re}A_{w}$ and $\operatorname{Im}A_{w}$ from the
mean pointer position. These sensitivities are the positive square roots of
the following expressions:%
\begin{equation}
\delta_{q}^{2}\operatorname{Re}A_{w}\equiv\frac{\Delta_{\Psi}^{2}q}{\left\vert
\frac{\partial\left\langle \Psi\right\vert \widehat{q}\left\vert
\Psi\right\rangle }{\partial\operatorname{Re}A_{w}}\right\vert ^{2}}%
=\frac{\Delta_{\phi}^{2}q}{\gamma^{2}}+\frac{2m}{3\gamma\hbar}%
\operatorname{Im}A_{w}\left(  \frac{dq_{3}}{dt}\right)  \label{S1}%
\end{equation}
(this quantity is obviously undefined when $A_{w}$ is purely imaginary) and%
\begin{equation}
\delta_{q}^{2}\operatorname{Im}A_{w}\equiv\frac{\Delta_{\Psi}^{2}q}{\left\vert
\frac{\partial\left\langle \Psi\right\vert \widehat{q}\left\vert
\Psi\right\rangle }{\partial\operatorname{Im}A_{w}}\right\vert ^{2}}=\left(
\frac{\hbar}{\gamma m}\right)  ^{2}\left(  \frac{\Delta_{\phi}^{2}%
q}{\left\vert \frac{d\Delta_{\phi}^{2}q}{dt}\right\vert ^{2}}\right)
+\frac{2}{3}\left(  \frac{\hbar}{\gamma m}\right)  \operatorname{Im}%
A_{w}\left(  \frac{\frac{dq_{3}}{dt}}{\left\vert \frac{d\Delta_{\phi}^{2}%
q}{dt}\right\vert ^{2}}\right)  ,\frac{d\Delta_{\phi}^{2}q}{dt}\neq0
\label{S2}%
\end{equation}
(this quantity is obviously undefined when $A_{w}$ is real valued or when
$\frac{d\Delta_{\phi}^{2}q}{dt}=0$ - in which case the mean position does not
depend upon $\operatorname{Im}A_{w}$). It is clear from eqs.(\ref{S1}) and
(\ref{S2}) that: (i) these measurement sensitivities depend upon the variance
control term $\operatorname{Im}A_{w}\left(  \frac{dq_{3}}{dt}\right)  $ and
that this dependence vanishes when $q_{3}$ is fixed (or if $A_{w}$ is real
valued); (ii) these measurement accuracies decrease (increase) as the
measurement gets weaker (stronger) - i.e. as $\gamma$ gets smaller (larger);
(iii) in principle - the accuracies associated with measuring both
$\operatorname{Re}A_{w}$ and $\operatorname{Im}A_{w}$ can be arbitrarily
increased (for a fixed $\gamma>0$ and $m$) by invoking condition (\ref{SQ})
and choosing $\operatorname{Im}A_{w}\left(  \frac{dq_{3}}{dt}\right)
=-\left(  \frac{3\hbar}{2\gamma m}\right)  \Delta_{\phi}^{2}q+\epsilon$, where
$\epsilon$ is a small positive real number; and (iv) surprisingly, the
measurement accuracy for $\operatorname{Re}A_{w}$ decreases with increasing
pointer mass, whereas that for $\operatorname{Im}A_{w}$ increases.

The sensitivity $\delta_{p}\operatorname{Im}A_{w}$ for determining
$\operatorname{Im}A_{w}$ from the mean pointer momentum is the positive square
root of%
\begin{equation}
\delta_{p}^{2}\operatorname{Im}A_{w}\equiv\frac{\Delta_{\Psi}^{2}p}{\left\vert
\frac{\partial\left\langle \Psi\right\vert \widehat{p}\left\vert
\Psi\right\rangle }{\partial\operatorname{Im}A_{w}}\right\vert ^{2}}=\left(
\frac{\hbar}{2\gamma}\right)  ^{2}\left(  \frac{1}{\Delta_{\phi}^{2}p}\right)
+\left(  \frac{\hbar}{2\gamma}\right)  \operatorname{Im}A_{3}\left(
\frac{p_{3}}{\left(  \Delta_{\phi}^{2}p\right)  ^{2}}\right)  ,\Delta_{\phi
}^{2}p\neq0 \label{S3}%
\end{equation}
(this quantity is undefined when $A_{w}$ is real valued). Inspection of
eq.(\ref{S3}) reveals that for such measurements: (i) the sensitivity depends
upon the variance control term $\operatorname{Im}A_{w}\left(  p_{3}\right)  $;
(ii) the accuracy decreases (increases) as the measurement gets weaker
(stronger) - i.e. as $\gamma$ gets smaller (larger); and (iii) the accuracy
can be arbitrarily increased (for a fixed $\gamma>0$) via eq.(\ref{SQA}) by
choosing $\operatorname{Im}A_{w}\left(  p_{3}\right)  =-\left(  \frac{\hbar
}{2\gamma}\right)  \Delta_{\phi}^{2}p+\epsilon$, where $\epsilon$ is again a
small positive real number.

In closing, it is important to note that the results discussed and developed
above apply when the measurement interaction is instantaneous and the
measurement is read from the pointer immediately after the interaction
\cite{DE}.

\section{Acknowledgement}

This work was supported in part by a grant from the Naval Surface Warfare
Center Dahlgren Division's In-house Laboratory Independent Research Program
sponsored by the Office of Naval Research.


\section{References}

\begin{thebibliography}{99}                                                                                               %


\bibitem {A1}Aharonov Y, Albert D, Casher D and Vaidman L 1986 \textit{New
Techniques and Ideas in Quantum Measurement Theory} ed D Greenberger (New
York: New York Academy of Sciences) p 417

\bibitem {A2}Aharonov Y, Albert D and Vaidman L 1988 \textit{Phys. Rev. Lett.}
\textbf{60} 1351

\bibitem {A3}Aharonov Y and Vaidman L 1990 \textit{Phys. Rev. A }\textbf{41} 11

\bibitem {R}Ritchie N, Storey J and Hulet R 1991 \textit{Phys. Rev. Lett.
}\textbf{66} 1107

\bibitem {P}Parks A, Cullin D and Stoudt D 1998 \textit{Proc. R. Soc.
A}\textbf{ 454} 2997

\bibitem {RL}Resch K, Lundeen J and Steinberg A 2004 \textit{Phys. Lett.
A}\textbf{ 324} 125

\bibitem {W}Wang Q, Sun F, Zhang Y, Li J, Huang Y and Guo G 2006 \textit{Phys.
Rev. A}\textbf{ 73} 023814

\bibitem {H}Hosten O and Kwiat P 2008 \textit{Science}\textbf{ 319} 787

\bibitem {Y}Yokota K, Yamamoto T, Koashi M and Imoto N 2009 \textit{N. J.
Phys.}\textbf{ 11} 033011

\bibitem {D}Dixon P, Starling D, Jordan A and Howell J 2009 \textit{Phys. Rev.
Lett.}\textbf{ 102} 173601

\bibitem {J}Johansen L 2004 \textit{Phys. Rev. Lett.} \textbf{93} 120402

\bibitem {AB}Aharonov Y and Botero A 2005 \textit{Phys. Rev. A}\textbf{ 72 }052111

\bibitem {Jo}Jozsa R 2007 \textit{Phys. Rev. A}\textbf{ 76 }044103

\bibitem {DE}Di Lorenzo A and Egues J 2008 \textit{Phys. Rev. A}\textbf{ 77} 042108

\bibitem {C}Cho Y, Lim H, Ra Y and Kim Y 2010 \textit{N. J. Phys.}\textbf{ 12} 023036

\bibitem {A4}Aharonov Y and Rohrlich D 2005 \textit{Quantum Paradoxes: Quantum
Theory for the Perplexed}$\mathbb{\ }$(Wiley-VCH, Weinheim) p.225
\end{thebibliography}
\end{document}